\documentclass[11pt]{amsart}
\usepackage{amsbsy,amssymb,amsmath,amsthm,amscd,amsfonts,latexsym,amstext,delarray,
amsmath,graphicx} 
\usepackage[margin=1in]{geometry}
\usepackage{color}
\input xypic

\newtheorem{thm}{Theorem}[section]
\newtheorem{prop}[thm]{Proposition}
\newtheorem{cor}[thm]{Corollary}
\newtheorem{lem}[thm]{Lemma}

\newtheorem{defn}[thm]{Definition}
\newtheorem{rem}[thm]{Remark}
\newtheorem{ex}[thm]{Example}

\numberwithin{equation}{section}

\def\Q{{\mathbb Q}}
\def\Z{{\mathbb Z}}
\def\N{{\mathbb N}}
\def\R{{\mathbb R}}
\def\C{{\mathbb C}}
\renewcommand\P{{\mathbb P}}

\def\P{{\mathbb P}}

\def\GL{{\rm GL}}

\def\cA{{\mathcal A}}

\def\cC{{\mathcal C}}

\def\cH{{\mathcal H}}

\def\cM{{\mathcal M}}

\def\cO{{\mathcal O}}

\def\cR{{\mathcal R}}
\def\cS{{\mathcal S}}

\def\bA{{\mathbb A}}

\def\bK{{\mathbb K}}

\DeclareMathOperator*{\Hom}{Hom}

\title{Quantum Computation and Real Multiplication}
\author{Matilde Marcolli and John Napp}
\address{Division of Mathematics, Physics and Astronomy, 
California Institute of Technology, 1200 E. California Blvd. Pasadena, CA 91125, USA}
\email{matilde@caltech.edu}
\email{jnapp@caltech.edu}
\date{}

\begin{document}

\begin{abstract}
We propose a construction of anyon systems associated to quantum tori with
real multiplication and the embedding of quantum tori in AF algebras.
These systems generalize the Fibonacci anyons, with weaker categorical properties,
and are obtained from the basic modules and the real multiplication structure.
\end{abstract}

\maketitle
%\tableofcontents

\section{Introduction}

Quantum computation refers, broadly, to the use of quantum phenomena such as entanglement and superposition to perform operations on data. It is believed that quantum computation is significantly more powerful than classical computation in certain respects, and could lead to advances in many areas of computation such as quantum simulation, cryptography, and database searching. In the quantum setting, one thinks of computation in terms of \emph{qubits} instead of classical bits. While classical bits take on precisely one of two possible states (e.g. $0$ or $1$), the state of a qubit may be in a superposition of two orthonormal states:
\[ |\psi \rangle = \alpha|0\rangle + \beta|1\rangle \, . \]
In the above expression, $\alpha$ and $\beta$ are complex numbers, and $|0\rangle$ and $|1\rangle$ form an orthonormal basis for this vector space. An example of an entangled state of two qubits is
\[ \beta_{00} \equiv \frac{1}{\sqrt{2}}(|00\rangle + |11\rangle)\, . \]
In the above state, measurement of one qubit results in a collapse of the wavefunction 
of the other qubit as well.

\smallskip

\emph{Quantum gates}, analogous to classical gates, are implemented by applying unitary operators to qubits. For example, consider the single-qubit $X$ and $Z$ gates, defined by
\[ X \equiv \left(\begin{array}{cc}0 & 1 \\ 1 & 0 \end{array}\right)\, , \ \ \ \ 
 Z \equiv \left(\begin{array}{cc}1 & 0 \\ 0 & -1 \end{array}\right) \, . \]
We see that $X$ can be interpreted as a bit flip (since $X|0\rangle = |1\rangle$) and $Z$ can be interpreted as a phase flip (since $Z|+\rangle=|-\rangle$, where $|\pm\rangle \equiv \frac{1}{\sqrt{2}}(|0\rangle \pm |1\rangle)$). Similarly, one can construct quantum gates that act on two or more qubits. 

\smallskip

A set of quantum gates is \emph{universal} if any operation possible on a quantum computer can be realized to arbitrary accuracy by a finite sequence of gates in the set.
A well known example of an anyon system that is universal for quantum computation is
known as the {\em Fibonacci anyons}, \cite{Hormo}, \cite{NSSFS}, \cite{Pres}. 

\smallskip

In this paper, we make a proposal for a possible construction of 
anyon systems based on the geometry of
quantum tori with real multiplication. The Fibonacci anyons are recovered
as a special case. Except in the Fibonacci case, however, the systems described here have weaker
properties from the categorical standpoint. In particular, they do not arise from
modular tensor categories. Nonetheless, they still exhibit some of the behavior of the
Fibonacci anyons. 

\smallskip

Our construction is based on the embedding of quantum tori in AF algebras and
on the description of AF algebras in terms of Bratteli diagrams, and on the basic
modules of the quantum torus that give the real multiplication structure when
the modulus is a quadratic irrationality. 

\smallskip

We first describe how anyon systems in the usual sense (those arising from
modular tensor categories and the associated fusion rules) determine
Bratteli diagrams describing the fusion paths and associated AF algebras
of operators acting on the Hilbert space of the system. In the Fibonacci
case, we show that this gives the usual description of the anyon system.
We also show that, in general, the $K_0$-group of the AF algebra arising
from an anyon system can be described in terms of the eigenvalues of
the fusion matrices and the number field generated by them. 

\smallskip

The opposite process, from AF algebras to anyon systems, cannot always
be carried out, but we focus on only those AF algebras $\bA_\theta$ in which quantum
tori $\cA_\theta$ with real multiplication embed, with the embedding inducing an
isomorphism on $K_0$. In this setting we use the geometry of the
quantum torus to provide a candidate for an associated anyon system.

\smallskip

We first show that the AF algebra $\bA_\theta$ can be equivalently described by
a Bratteli diagram with incidence matrix given by a matrix in $\GL_2(\Z)$ that
fixes $\theta$ (which exists because $\theta$ is a quadratic irrationality), or by
a Bratteli diagram with incidence matrix given by the fusion matrix $N_1$ of the
anyon system, of the form
$$ N_1= \left( \begin{array}{cc} {\rm Tr}(g) & 1 \\ 1 & 0 \end{array}\right), $$ 
where $x_0={\bf 1}$ and $x_1$ are the two anyon types in the system, 
with $x_0\otimes x_i=x_i\otimes x_0=x_i$ and $x_1\otimes x_1 = x_1^{\oplus {\rm Tr}(g)} \oplus x_0$.
The Fibonacci case corresponds to ${\rm Tr}(g)=1$. We compute explicitly the $S$-matrix of the
resulting fusion rule. The approximations in the AF algebra $\bA_\theta$ to 
the generators $U$ and $V$ of the quantum torus are phase shifter and downshift permutation
quantum gates on the Hilbert space of the fusion paths of the anyon system.

\smallskip

We show that the fusion ring determined by the fusion rules of this anyon system
is isomorphic to $K_0(\cA_\theta)$ with the product given by identifying it, via
the range of the trace, with an order in the real quadratic field $\Q(\theta)$.
We then propose that a categorification of this fusion ring is obtained by
considering basic modules $E_{g^k}(\theta)$, $k\in \Z$, with the tensor
product $E_g(\theta) \otimes_{\cA_\theta} \cdots \otimes_{\cA_\theta} E_g(\theta)
=E_{g^k}(\theta)$. Thus, the basic module $X_1=E_g(\theta)$ with $g(\theta)=\theta$
and $X_0=\cA_\theta$ have classes $x_1=[E_g(\theta)]$ and $x_0=1$ in $K_0(\cA_\theta)$
that satisfy the fusion rules. {}From this perspective, we describe the $F$-matrices as
homomorphisms between sums of basic modules on the quantum torus, and we
formulate a version of the pentagon relation as an equation in the quantum torus $\cA_\theta$.
We suggest a possible approach to constructing solutions using elements in the algebra
that satisfy a pentagon identity, related to the quantum dilogarithm function.

\smallskip

In the original Fibonacci case, we reformulate the braiding action in terms of
the AF algebra and we show that elements of the $K_0$-group, seen as 
dimension functions, determine associated disconnected braidings of the anyons.

\bigskip
\section{Anyon systems}

An anyon is a 2-dimensional quasiparticle with nontrivial braiding statistics, that is, 
swapping two identical anyons induces a nontrivial phase shift of the wavefunction. For anyons with abelian braiding statistics:
\begin{equation}\label{phaseshift}
\psi \mapsto e^{\pi i \theta} \psi\, , \ \ \ \text{ with } \ \ \  \theta\in \mathbb{R} \setminus \mathbb{Z}. 
\end{equation}
More generally, one may have anyons with non-abelian braiding statistics, corresponding to higher-dimensional representations of the braid group: $$\psi_\alpha \mapsto \sum_\beta \rho_{\alpha \beta} \psi_\beta,$$ with $\rho$ a unitary square matrix with dimension corresponding to the degeneracy of the system, see \cite{NSSFS}.

Note that this property is unique to two dimensions; in three spacial dimensions, swapping two particles results in a phase shift of only $+1$ (bosons) or $-1$ (fermions). One performs computation by swapping these anyons in various ways, which can be interpreted as acting on the system by the braid group. The result of a computation is measured by fusing the anyons together in a certain order, and measuring the topological $q$-spin of the resulting product at each step of the fusion. The set of possible fusion paths is the basis for the corresponding Hilbert space of the system.

\smallskip
\subsection{Fusion rules}

An anyon system is specified by a list of the different particle types and fusion rules
that assign to a pair of anyons a combination of resulting anyons, 
\begin{equation}\label{fusion}
x_i \otimes x_j = \oplus_k N^k_{ij} x_k, 
\end{equation}
where the non-negative integers $N_{ij}^k$ specify the
admissible fusion channels. 

\smallskip

One requires that one of the anyon types corresponds to the vacuum ${\bf 1}$, 
the ground state of the system that satisfies
$$ {\bf 1} \otimes {\bf 1} = {\bf 1} \ \ \ \text{ and } \ \ \ 
{\bf 1} \otimes x_i = x_i \otimes {\bf 1} = x_i $$
for all anyon types in the system. We will label the anyon types $\{ x_0, \ldots, x_N \}$
so that $x_0 ={\bf 1}$.

\smallskip

One also assumes that each anyon $x_i$ has a dual $x_i^*$, which is also
one of the anyons in the list. The vacuum ${\bf 1}$ is self-dual.

\smallskip

The fusion coefficients satisfy the identities
$$ N_{0j}^k=\delta_{jk}, \ \ \ N^0_{ij} =\delta_{ij'} $$
and
$$ N^k_{ij}=N^k_{ji} = N^{j'}_{i k'}=N^{k'}_{i'j'}, $$
where $x_{i'}=x_i^*$ denotes the dual anyon, 
so that, if $N_i$ is the matrix with $(N_i)_{jk}=N^k_{ij}$,
then $N_{i'}=N_i^t$, the transpose matrix, \cite{Row}.

\smallskip

Let $\Lambda_i={\rm diag}( \lambda_{ij} )$ be the diagonal matrix 
with entries the eigenvalues $\lambda_{ij}$ of the matrix $N_i$,
and let $S_i$ be the matrix with columns the corresponding
eigenvectors.

\smallskip
\subsection{Categorical setting}

We recall briefly some categorical notions relevant to anyon based
quantum computing. For more details, we refer the reader to \cite{Eting}
and other references below.

\smallskip

Modular tensor categories provide the typical categorical setting considered 
 to describe anyon systems (see \cite{Kit}, \cite{Wang}). However, here we
 will be considering systems that satisfy weaker categorical structures,
 hence we review here the simplest levels of categorical structure that
 we will be working with.
 
 \smallskip

A category $\cC$ is {\em semi-monoidal} if it is endowed  with a functor
$\otimes: \cC \times \cC \to \cC$, and with natural  associativity isomorphisms
$$ \tau_{X,Y,Z}: X\otimes(Y\otimes Z) \to (X\otimes Y)\otimes Z $$
for all $X,Y,Z\in {\rm Obj}(\cC)$ satisfying the pentagon relation
$$ (\tau_{X,Y,Z}\otimes 1_W) \circ \tau_{X,Y\otimes Z, W} \circ (1_X \otimes \tau_{Y,Z,W})
= \tau_{X\otimes Y,Z,W}\circ \tau_{X,Y,Z\otimes W}. $$
It is {\em braided semi-monoidal} if it also has natural symmetry isomorphisms
$$ \sigma_{X,Y}: X\otimes Y\to Y\otimes X $$
for all $X,Y \in {\rm Obj}(\cC)$ satisfying the hexagon relation
$$ \tau_{X,Y,Z} \circ \sigma_{X\otimes Y,Z} \circ \tau_{X,Y,Z} =
(\sigma_{X,Z}\otimes 1_Y) \circ \tau_{X,Z,Y} \circ (1_X \otimes \sigma_{Y,Z}). $$
The braiding is symmetric if $\sigma_{X,Y}\circ \sigma_{Y,X} =1_X$ for all
 $X,Y \in {\rm Obj}(\cC)$. 
 
 \smallskip
 
 A category $\cC$ is {\em monoidal} (or tensor) if it is semi-monoidal and has a
 unit object ${\bf 1} \in {\rm Obj}(\cC)$ with natural isomorphisms
 $\lambda_X : {\bf 1}\otimes X \to X$ and $\rho_X: X \otimes {\bf 1} \to X$
 satisfying the triangle relation 
 $$ 1_X \otimes \lambda_Y = (\rho_X \otimes 1_Y) \circ \tau_{X,{\bf 1}, Y}. $$
 
 \smallskip
 
 Given a semi-monoidal category $\cC$, its unit augmentation $\cC^+$ is
 a monoidal category given by the coproduct $\cC\amalg I$, where $I$ is
 the trivial group seen as a category with a single object ${\bf 1}$
 and morphism $1_{\bf 1}$, and with the tensor product  $\otimes$
 on $\cC$ extended by $-\otimes {\bf 1} ={\bf 1}\otimes - =Id_{\cC\amalg I}$,
 see \cite{Hines}.
 
 \smallskip
 
 A category $\cC$ is {\em braided monoidal} if it is monoidal and braided
 semi-monoidal, with the compatibility between the braiding and the unit
 given by 
 $$ \lambda_X \circ \sigma_{X,{\bf 1}} = \rho_X. $$
 
 \smallskip
 
 A category $\cC$ is $\C$-{\em linear} if $\Hom_\cC(X,Y)$ are $\C$-vector spaces,
 for all $X,Y\in {\rm Obj}(\cC)$. It is {\em semisimple} if it is an abelian category
 in which every object can be written as a direct sum of simple objects, and {\em finite} 
 if there are only finitely 
 many simple objects up to isomorphism. If $\cC$ is a monoidal category and it
 is $\C$-linear finite semisimple, one assumes also that the unit object ${\bf 1}$
 is a simple object. 
 A {\em fusion category} is a $\C$-linear finite semisimple rigid tensor category
where all the $\Hom$ spaces are finite dimensional.
 
\smallskip
\subsection{Fusion rules and categorification}

Specifying anyon systems on the basis of the particle types and fusion rules
leads to a set of discrete data, which can be expressed algebraically in the
form of a fusion ring. The underlying categorical setting can then be seen as
a categorification of the fusion rules and fusion ring.

\smallskip

We can formulate more precisely this procedure as in \cite{RoStoWa}, in the
form of {\em modular data}, which may or may not come from a modular
tensor category.

\smallskip

\begin{defn}\label{ModDataDef}
A {\rm modular fusion rule} consists of 
\begin{enumerate}
\item a set $\{ x_i \}_{i=1,\ldots, N}$ of anyon types, 
\item a charge conjugation matrix $C_{ij}=\delta_{ii'}$ where $x_{i'}=x_i^*$
is the dual,
\item a set of $N\times N$-matrices $N_i=(N_{ij}^k)$, for $i=1,\ldots, N$ with entries
in $\Z_{\geq 0}$ and with a common basis of eigenvectors,
\begin{equation}\label{NSi}
N_i = \tilde S \, \Lambda_i \, \tilde S^\dagger,
\end{equation}
where $\Lambda_i={\rm diag}(\lambda_{ij})$ is the diagonal matrix with
entries the eigenvalues of $N_i$ and the columns of $\tilde S$ are the corresponding 
eigenvectors.
\end{enumerate}
\end{defn}

{\em Modular fusion data} (see Definition 2.1 of \cite{RoStoWa}) satisfy the
properties above and are additionally endowed with 
a diagonal matrix $T$ with $T_{ij}=\delta_{ij} \theta_j$ and $\theta_i \in U(1)$,
satisfying $(ST)^3= D_+ S_{00}\, S^2$, where $S=D^{-1} \tilde S$ with $D=(\sum_i d_i^2)^{1/2}$
with $d_i$ the Perron--Frobenius eigenvalue of $N_i$, and $D_+=\sum_i \theta_i d_i^2$.
Not all modular fusion data come from modular tensor categories, see the classification
results of \cite{RoStoWa}. We will not be concerned with modular tensor categories here,
and we will only look at a weaker requirement for a categorification of fusion rules.

\smallskip

The {\em fusion ring} $\cR$ associated to a modular fusion rule is obtained by considering
the polynomial ring $\Z[x_1, \ldots, x_N]$ modulo the ideal generated by the relations
$x_i x_j = \sum_k N^k_{ij} x_k$.  The categorification problem then consists of constructing
a braided monoidal category $\cC$ with $K_0(\cC)=\cR$.

\smallskip
\subsection{Fibonacci anyons}

A simple but very interesting model of anyon system used in topological quantum computation
is known as the {\em Fibonacci anyon model} (\cite{Hormo}, \cite{NSSFS}, \cite{Pres}).

\smallskip

In this model we have two possible particles, denoted by $x_0={\bf 1}$ and $x_1=\tau$. 
The fusion rules are 
\begin{equation}\label{Fibfusion}
x_0\otimes x_0=x_0, \ \ \  x_0\otimes x_1=x_1\otimes x_0=x_1, \ \ \ \text{ and } \ \ \  x_1\otimes x_1=x_0\oplus x_1. 
\end{equation}
The final fusion rule above implies that fusing two $q$-spin 1 quasiparticles results in a quasiparticle with $q$-spin in a superposition of $0$ and $1$. This model is enticing because it is the simplest topological model that is universal, in the sense that every unitary operation on the Hilbert 
space can be approximated to arbitrary accuracy by braiding alone. 

\smallskip

Our goal is to reformulate and study this system and some direct generalizations in noncommutative geometry terms.

\section{Anyon systems and AF algebras}

We show that fusion rules for anyon systems determine the data of a stationary Bratteli
diagram, which specifies an AF (approximately finite dimensional) $C^*$-algebra
with a representation on a Hilbert space that is isomorphic to the state space of the
anyon system described in terms of fusion paths. We also show that the $K_0$
group of the AF algebra, which can be computed as a direct limit of abelian groups
over the Bratteli diagram, has an explicit expression in terms of the quantum dimensions
of the anyon system and it is endowed with a product structure induced by the fusion
ring of the anyon system. 

\smallskip
\subsection{Bratteli diagrams and AF algebras}
We recall here a few well known facts about Bratteli diagrams and AF algebras
that we need to use in the following (see \cite{Bra}, \cite{Ell}).

\smallskip

A Bratteli diagram is an infinite directed graph $\Gamma=(V,E)$ where the vertex 
set has a partitioning $V=\cup_{n\geq 0} V_n$ and the edge set has a partitioning
$E=\cup_{n\geq 0} E_n$, where $E_n$ is the set of oriented edges 
with source vertex in $V_n$ and target vertex in $V_{n+1}$. One further
assumes that $V_0$ consists of a single vertex. The incidence matrices 
$\varphi_n$ of the Bratteli diagrams are given by $(\varphi_n)_{ij}=\# \{ e \in E_n\,|\,
s(e)=v_i \in V_n, \, t(e)=v_j \in V_{n+1} \}$.

\smallskip

Given a Bratteli diagram $\Gamma$ one can form new diagrams $\Gamma'$ 
by {\em telescoping}. Namely, given a sequence $0< m_1< \cdots < m_k < \cdots$,
the new diagram $\Gamma'$ has $V'_n =V_{m_n}$ and edge set $E'_n$
given by all the possible directed paths in the original diagram $\Gamma$
with source in $V_{m_n}$ and target in $V_{m_{n+1}}$. The incidence
matrices are given by the product $\varphi'_n = \prod_{k=m_n}^{m_{n+1}-1} \varphi_k$. 

\smallskip

An AF algebra $\bA$ is a $C^*$-algebra given by a direct limit 
$$ \bA = \varinjlim_n \bA_n $$
of finite dimensional 
$C^*$-algebras $\bA_n$. By Wedderburn's theorem, the latter are direct sums of
matrix algebras over $\C$, 
$$ \bA_n = \oplus_{k=1}^{N_n} \cM_{r_k}(\C). $$
An AF algebra can therefore always be described in
terms of a Bratteli diagram (see \cite{Bra}), where one has $\# V_n =N_n$ with
the vertices in $V_n$ decorated by the direct summands $\cM_{r_k}(\C)$.
The embedding $\bA_n \hookrightarrow \bA_{n+1}$ of the direct system of
finite dimensional algebras is determined by the incidence matrix $\varphi_n$,
namely, one embeds the matrix algebra $\cM_{r_i}(\C)$ attached to the vertex
$v_i \in V_n$ in the matrix algebra $\cM_{r_j}(\C)$ attached to the vertex
$v_j \in V_{n+1}$ with multiplicity $(\varphi_n)_{ij}$,
$$ \cM_{r_i}(\C)^{\oplus (\varphi_n)_{ij}} \hookrightarrow \cM_{r_j}(\C). $$

\smallskip

Isomorphism of Bratteli diagrams $\Gamma=(V,E)$ and $\Gamma'=(V',E')$ 
is defined as a bijection of $V$ and $V'$ and $E$ and $E'$ preserving the
grading, that intertwines the source and target maps.

\smallskip

One considers the set of Bratteli diagrams up to the equivalence relation generated by
isomorphisms and telescoping. It is shown in \cite{Bra} that the equivalence class of the
Bratteli diagram under this equivalence relation is a complete isomorphism
invariant of the AF algebra. A complete isomorphism invariant of AF algebras of
$K$-theoretic nature is given by the dimension group (see \cite{Ell}), which is the
ordered $K_0$-group of the AF algebra and can be obtained as a direct limit of
abelian groups associated to the Bratteli diagram.

\smallskip

A Bratteli diagram $\Gamma$ is {\em stationary} if $\# V_n =N$ and $\varphi_n=\varphi$ 
for all $n\geq 1$. It is {\em simple} if there is a telescoping $\Gamma'$ such that all entries 
of the incidence matrices $\varphi_n'$ are positive.

\smallskip

Associated to a Bratteli $\Gamma$, we also consider a 
Hilbert space $\cH$, obtained as a direct limit of the system
of finite dimensional Hilbert spaces 
\begin{equation}\label{Hn}
\cH_n =\oplus_{k=1}^{N_n} \C^{r_k}
\end{equation}
with embeddings $\cH_n \hookrightarrow \cH_{n+1}$ determined by
the incidence matrices $\varphi_n$ of $\Gamma$, with suitable normalization
factors so that the embeddings are isometries, as in (2.32) -- (2.35) of \cite{LaLiSza}.

\smallskip

\subsection{Fibonacci AF algebra and Fibonacci anyons}

We reformulate the fusion properties of a system of Fibonacci anyons in
terms of the data of an AF algebra, determined via a Bratteli diagram.

\begin{prop}\label{propBrat}
The fusion properties of a system of Fibonacci anyons with an arbitrarily large 
number of particles are encoded in the Fibonacci AF $C^*$ algebra given by
the direct limit $\bA=\varinjlim_n \bA_n$ of matrix algebras 
$\bA_n=\mathcal{M}_{\textrm{Fib}(n)}(\C)\oplus \mathcal{M}_{\textrm{Fib}(n-1)}(\C)$,
with $\textrm{Fib}(n)$ the $n$-th Fibonacci number, with embeddings 
$\phi_n: \bA_n \hookrightarrow \bA_{n+1}$ implemented by the matrix 
describing the basic fusion rules \eqref{Fibfusion},
\begin{equation}\label{Fibmatrix}
\varphi=\left(\begin{array}{cc}1 & 1 \\ 1 & 0 \end{array}\right)
\end{equation}
\end{prop}

\proof
One can diagrammatically describe the fusion properties of configurations anyons via a Bratteli diagram, which depicts the possible fusion paths in a system of an arbitrary number of particles. The Bratteli diagram for a system of Fibonacci anyons is of the form
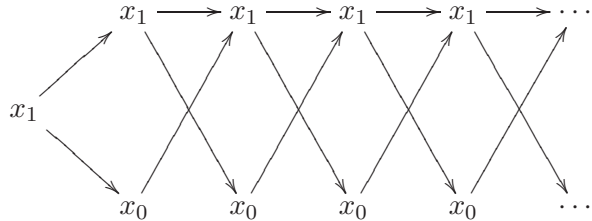
\[\xymatrix{
& x_1 \ar[r] \ar[rdd] & x_1 \ar[r] \ar[rdd] & x_1 \ar[r] \ar[rdd] & x_1 \ar[r] \ar[rdd] & \cdots \\
x_1 \ar[ur] \ar[dr] \\
& x_0 \ar[ruu]  & x_0 \ar[ruu] & x_0 \ar[ruu] & x_0 \ar[ruu] & \cdots}  \]
where $x_1$ represents a Fibonacci anyon and $x_0={\bf 1}$ represents the particle with trivial braiding statistics. Note that this diagram depicts how an $x_1$ particle fused with an $x_1$ particle can result in an $x_0$ or $x_1$ particle, respectively, with probabilities $p(x_0)=\tau^{-2}$ and $p(x_1)=\tau^{-1}$,
with $\tau=(1+\sqrt{5})/2$ the golden ratio (see \cite{Treb}); while an $x_0$ fused with an $x_1$ can result only in an $x_1$. For such a system with $N$ anyons, there will be $N$ columns in the corresponding Bratteli diagram.

\smallskip

More precisely, we check that the Hilbert space of the system, as defined by the Bratteli
diagram, agrees with the Hilbert space of the anyon system.

\smallskip

Recall that a basis of the Hilbert space of the system is given by the set of fusion paths.
It is known that the fusion tree basis for fusions of $N$ Fibonacci anyons, that is the
counting of all possible labelings of the fusion paths, has cardinality  $\textrm{Fib}(N)$,
see \cite{Treb}.

\smallskip

Thus, we can identify the Hilbert space $\cH_N=\C^{\textrm{Fib}(N)}$ 
of a system of $N$ Fibonacci anyons system with the Hilbert space 
$\C^{\textrm{Fib}(N)} \simeq \C^{\textrm{Fib}(N-1)} \oplus \C^{\textrm{Fib}(N-2)}$
constructed after $N$ steps of the Bratteli diagram above, where we 
replace the labeling $x_0$ and $x_1$ 
on the nodes with the dimensions of the corresponding Hilbert space 
at each level and total $q$-spin. The diagram is given by 
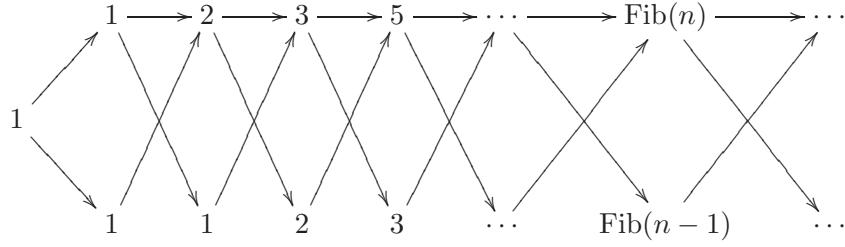
\[\xymatrix{
& 1 \ar[r] \ar[rdd] & 2 \ar[r] \ar[rdd] & 3 \ar[r] \ar[rdd] & 5 \ar[r] \ar[rdd] & \cdots \ar[r]\ar[rdd] & \textrm{Fib}(n)\ar[r]\ar[rdd] & \cdots \\
1 \ar[ur] \ar[dr] \\
& 1 \ar[ruu]  & 1 \ar[ruu] & 2 \ar[ruu] & 3 \ar[ruu] & \cdots \ar[ruu] & \textrm{Fib}(n-1)\ar[ruu] & \cdots} \]

\smallskip

We recognize this diagram as describing the direct system of Hilbert spaces $\cH_N$
that give a representation of the Fibonacci AF algebra (see \cite{Dav}). This is the AF $C^*$-algebra 
$\bA$ given by $\bA=\overline{\cup_{n\geq 1}\bA_n}$ with $\bA_n=\mathcal{M}_{\textrm{Fib}(n)}(\C)\oplus \mathcal{M}_{\textrm{Fib}(n-1)}(\C)$ where $\mathcal{M}_n(\C)$ is the full matrix algebra of $n\times n$ matrices, and the embeddings of $\bA_n$ in $\bA_{n+1}$ are as specified in the diagram.
The AF algebra $\bA$ is obtained as the direct limit over the stationary Bratteli diagram with
$$ \varphi_n = \left( \begin{array}{cc} 1 & 1 \\ 1 & 0 \end{array}\right). $$ 
In terms of the anyons model, the Fibonacci AF algebra $\bA=\bA_\tau$ 
is the algebra of operators acting on the Hilbert space of the anyon system 
given by a direct limit $\cH=\varinjlim_n \cH_n$ with $\dim \cH_n = \textrm{Fib}(n+1)=
\textrm{Fib}(n)+\textrm{Fib}(n-1)$.
\endproof

\smallskip

\begin{cor}\label{Fibflip}
The Fibonacci
AF algebra computed by the Bratteli diagram of Proposition \ref{propBrat} is isomorphic
to the AF algebra computed by the stationary Bratteli diagram with incidence matrix
\begin{equation}\label{Fibmatrix2}
\varphi=\left( \begin{array}{cc} 0 & 1 \\ 1 & 1 \end{array}\right).
\end{equation}
\end{cor}

\proof The incidence matrices \eqref{Fibmatrix} and \eqref{Fibmatrix2} are related
by a unitary conjugation
$$  \left( \begin{array}{cc} 0 & 1 \\ 1 & 0 \end{array}\right)
\left( \begin{array}{cc} 0 & 1 \\ 1 & 1 \end{array}\right)
 \left( \begin{array}{cc} 0 & 1 \\ 1 & 0 \end{array}\right) = \left( \begin{array}{cc} 1 & 1 \\ 1 & 0 \end{array}\right). $$
\endproof

\smallskip
\subsection{AF algebras from anyon systems}

Suppose given an anyon system, with fusion rules defined by the data
$\{ x_i \}_{i=0,\ldots, N-1}$ and $N^k_{ij}$ as in Definition \ref{ModDataDef}. For each $i$,
form a stationary Bratteli diagram $\Gamma_i$ with $\# V_n=N$ for all $n\geq 1$, with the single 
vertex of $V_0$ labelled by $x_0$ and the $N$ vertices of $V_n$ labelled by
$\{ x_i \}_{i=0,\ldots, N-1}$. The incidence matrices are taken to be 
$\varphi_{i,n} =\varphi_i= N_i$.

These Bratteli diagrams define AF algebras $\bA_i =\varinjlim_n \bA_{i,n}$ acting on
Hilbert spaces $\cH_i = \varinjlim_n \cH_{i,n}$, as in \eqref{Hn}, with 
$\bA_{i,n}= \cM_{d_{i,n}}(\C)^{\oplus N}$. 

\smallskip

\begin{lem}\label{NiAHanyon}
The Hilbert space $\cH=\oplus_i \cH_i$, with the $\cH_i$ determined by the Bratteli
diagram $\Gamma_i$ as above, is isomorphic to the Hilbert space of the anyon
system.  
\end{lem}

\proof The embeddings $\cH_{i,n} \hookrightarrow \cH_{i,n+1}$ are determined by
the incidence matrices $\varphi_{i,n}=\varphi_i$ of $\Gamma_i$, normalized so as to obtain
an isometry, see \cite{LaLiSza}. Thus, the Hilbert space 
$\cH_{i,n} =\oplus_{j=0}^{N-1}\cH_{i,n,j}$ has dimension
\begin{equation}\label{dindimHn}
d_{i,n}= \dim \cH_{i,n}= \sum_{j_1, \ldots, j_n} N_{ij_1}^{j_2} \cdots N_{ij_n}^{j_{n-1}} ,
\end{equation}
with the individual component $\cH_{i,n,j}$ of dimension 
$$ d_{i,n,j}= \dim \cH_{i,n,j}=\sum_{j_1, \ldots, j_{n-1}} N_{ij_1}^{j_2} \cdots N_{ij}^{j_{n-1}}. $$
The total Hilbert space $\cH$ is then given by the direct limit of the $\oplus_i \cH_{i,n}$ of
dimensions $d_n = \sum_i d_{i,n}$.
This counting of dimensions agrees with the usual counting of dimensions for the
Hilbert space of anyon fusion paths (see for instance \S 4.1.3 of \cite{Pachos}). 
\endproof

The AF algebra $\cA=\oplus_i \bA_i$ then acts as operators on the Hilbert
space $\cH$ of the anyon system. 

\smallskip

Two torsion-free abelian groups of finite rank $G_1$ and $G_2$ are quasi-isomorphic
if there exist morphisms $f_1: G_1 \to G_2$ and $f_2: G_2 \to G_1$ with $f_1\circ f_2= n \, id_{G_2}$
and $f_2\circ  f_1 = n\, id_{G_1}$ for some $n\in \N$.

\smallskip

For an algebraic integer $\lambda$ let $\bK_\lambda=\Q[\lambda]$ 
with $\cO_\lambda$ the ring of integers and set, as in \cite{Dugas},
\begin{equation}\label{Llambda}
L_\lambda = \cO_\lambda[\lambda^{-1}].
\end{equation}

\smallskip

We then obtain the following result relating the AF algebra to the number
field generated by the eigenvalues of the fusion matrices. For a detailed
study of the role of this number field in the setting of modular tensor
categories, we refer the reader to \cite{BCLDB}.

\smallskip

\begin{prop}\label{K0lambda}
The $K_0$-group $K_0(\bA_i)$ of the AF algebra $\bA_i$ of the anyon system is
quasi-isomorphic to 
\begin{equation}\label{Lijgroup}
L_{\bA_i}:= \bigoplus_j L_{\lambda_{ij}}, 
\end{equation}
where $\lambda_{ij}$ are the
eigenvalues of the fusion matrix $N_i$.
\end{prop}

\proof
The dimension group of an AF algebra $\bA$, that is, the $K_0$-group as a scaled ordered
group, can be computed as a direct limit over the Bratteli diagram, \cite{Ell}, with
\begin{equation}\label{K0lim}
(K_0(\bA), K_0(\bA)^+, [1_{\bA}])= \varinjlim_{\varphi_n} (K_0(\bA_n), K_0(\bA_n)^+, [1_{\bA_n}]). 
\end{equation}

\smallskip

For the stationary Bratteli diagram $\Gamma_i$ with $\varphi_{i,n}=\varphi_i$,
and with $\bA_{i,n}=\oplus_{j=0}^{N-1} \cM_{d_{i,n,j}}(\C)$, we have 
$K_0(\bA_{i,n})=\Z^N$, with the standard ordering and unit, and the limit is 
\begin{equation}\label{K0limAi}
(K_0(\bA_i), K_0(\bA_i)^+, [1_{\bA_i}])= \varinjlim_{\varphi_i} \Z^N. 
\end{equation}

\smallskip

We use the approach of \cite{Dugas} to evaluate direct limits of
the form \eqref{K0limAi}. Under the assumption that $\varphi: \Z^N \to \Z^N$
is an injective homomorphism, let ${\rm Spec}(\varphi)$ the set of eigenvalues
of $\varphi$. These are algebraic integers. It is shown in \cite{Dugas} that 
the direct limit is quasi-isomorphic to 
\begin{equation}\label{dilimlambda}
 G = \varinjlim_\varphi \Z^N \simeq 
 \bigoplus_{\lambda \in {\rm Spec}(\varphi)}  L_{\lambda}^{\oplus n_\lambda},
\end{equation}
where $n_\lambda$ is the dimension of the Jordan block $J_{n_\lambda}(\lambda)$ in the
Jordan normal form of $\varphi$.

\smallskip

The matrix $\varphi_i=N_i$ is by construction equal to 
$N_i = \tilde S \, \Lambda_i \, \tilde S^\dagger$, as in \eqref{NSi}
with $\Lambda_i={\rm diag}(\lambda_{ij})$ the diagonal matrix of eigenvalues. 
Thus, using the above, we can identify, up to quasi-isomorphism, the 
direct limit abelian group $K_0(\bA_i)$ with $\bigoplus_j L_{\lambda_{ij}}$. 
\endproof

Notice that $L_{\bA_i}=\bigoplus_j L_{\lambda_{ij}}$ above is just considered as
an abelian group. We now show that there is a ring structure on $\oplus_i L_{\bA_i}$
coming from the fact that the eigenvalues $\lambda_{ij}$ of the fusion matrices 
satisfy a Verlinde formula, and we compare it with the fusion ring of the anyon system.

\smallskip

\begin{lem}\label{LAring}
Consider the abelian group $L_{\bA}=\oplus_i L_{\bA_i}$ with $L_{\bA_i}$ as in \eqref{Lijgroup}.
The Verlinde formula for the eigenvalues $\lambda_{ij}$ of the fusion matrices
$N_i$ implies that $L_{\bA}$ has a ring structure that recovers the
fusion ring of the anyon system.
\end{lem}

\proof 
The eigenvalues $\lambda_{ij}$ of the fusion matrices $N_i$ satisfy the 
Verlinde formula (see in (8) of \cite{BCLDB})
$$ \lambda_{aj} \lambda_{bj} = \sum_c N^c_{ab} \lambda_{c,j}. $$
For a fixed $j$, this is the same multiplicative structure as in the fusion ring
of the anyon system $\Z[x_0,\ldots,x_{N-1}]/( x_i x_j -\sum_k N^k_{ij} x_k)$.
\endproof

\medskip

In general, one cannot conversely construct an anyon system from any
arbitrary data of an AF algebra and its Bratteli diagrams. However, we
will see in the next section that there is a particular class of AF algebras
in which quantum   tori with real multiplication embed with isomorphic
$K_0$, for which one can use the additional structure of the quantum  
torus with real multiplication to construct an anyon system. The Fibonacci
anyons are recovered as a particular case in this class.

\medskip
\section{Anyon systems and quantum   tori with real multiplication}

We now restrict our attention to AF algebras $\bA_\theta$ associated
to the continued fraction expansion of an irrational number $\theta$
that is a quadratic irrationality. It is well known \cite{PiVo} that the
quantum torus $\cA_\theta$ of modulus $\theta$ can be embedded
in the AF algebra $\bA_\theta$ in such a way as to induce an isomorphism
on $K_0$. We use the geometry of quantum tori with real multiplication
to construct anyon systems associated to an arbitrary quadratic 
irrationality $\theta$, which generalize the case of the Fibonacci
anyons, where $\theta=\tau = (1+\sqrt{5})/2$ is the golden ratio.
The real multiplication structure determines the fusion rules for
the resulting anyon system.

\smallskip
\subsection{AF algebras and quantum   tori}

The quantum   torus of modulus $\theta\in \R$ is the universal $C^*$-algebra $\cA_\theta$
generated by two unitaries $U,V$ satisfying the commutation relation
\begin{equation}\label{NCtorus}
UV = e^{2\pi i \theta} VU.
\end{equation}
We will be concerned with the case where the modulus is 
irrational $\theta\in \R\smallsetminus \Q$, and in particular .

\smallskip

It was shown in \cite{PiVo}, \cite{PiVo2} that the $C^*$-algebra $\cA_\theta$ 
of the quantum   torus can always be embedded into an AF algebra 
$\bA_\theta$ so that the embedding determines an isomorphism on $K_0$
preserving the positive cone. 
The AF algebra $\bA_\theta$ of \cite{PiVo} is obtained by considering the
continued fraction expansion of $\theta=[c_0, c_1, c_2, \ldots]$, with $q_n$ 
the denominators of the successive convergents of the expansion. One considers the algebras
\begin{equation}\label{contfrAn}
\bA_{n,\theta}= \cM_{q_n}(\C) \oplus \cM_{q_{n-1}}(\C).
\end{equation}
The embeddings $\bA_{n-1,\theta} \hookrightarrow \bA_{n,\theta}$ are given by
$$ \varphi_n = \left( \begin{array}{cc} 0 & 1 \\ 1 & c_n \end{array}\right), $$
$$ \cM_{q_{n-1}}(\C) \oplus \cM_{q_{n-2}}(\C) \to \cM_{q_{n-1}}(\C) \oplus 
\left( \cM_{q_{n-2}}(\C) \oplus \cM_{q_{n-1}}(\C)^{\oplus c_n}  \right)    \subset \cM_{q_n}(\C) , $$
where $c_n$ is the $n$-th digit of the continued fraction expansion, satisfying
$$ q_n = c_n q_{n-1} + q_{n-2}. $$

\smallskip

As in Corollary \ref{Fibflip}, one can equivalently work (as in \cite{LaLiSza}, \cite{PiVo})
with embeddings of the form
$$ \varphi_n = \left( \begin{array}{cc} c_n & 1 \\ 1 & 0 \end{array}\right). $$

\smallskip

Note that the AF algebra $\bA_\theta$ is not the quantum   torus itself. Rather, the 
quantum   torus $\cA_\theta$ is embedded into $\bA_\theta$. 

\smallskip
\subsection{Quantum tori with real multiplication}

We are especially interested here in the case of quantum   tori $\cA_\theta$
with irrational modulus $\theta \in \R\smallsetminus \Q$ that is a quadratic
irrationality, that is, a solution of a quadratic polynomial equation over $\Q$, or
equivalently $\theta \in \bK \smallsetminus \Q$, with $\bK$ a real quadratic field.
A theory of quantum   tori with real multiplication was initiated by Manin,
\cite{Manin}, as candidate geometric objects that may play, in the case of real
quadratic field, a similar role as elliptic curves with complex
multiplication play in the imaginary quadratic case. The theory was further
developed in \cite{Polishchuk} and \cite{Plazas}, \cite{Vlasenko}.

\smallskip

As in \cite{LewZag}, we denote by ${\rm Red}=\cup_{n\geq 1} {\rm Red}_n$ the 
semigroup of {\em reduced matrices} in $\GL_2(\Z)$, with
\begin{equation}\label{Redn}
{\rm Red}_n =\left\{ \left(\begin{array}{cc} 0 & 1 \\ 1 & k_1 \end{array}\right) \cdots
\left(\begin{array}{cc} 0 & 1 \\ 1 & k_n \end{array}\right) \,|\, k_1, \ldots, k_n \in \N \right\}.
\end{equation}
All reduced matrices are hyperbolic and 
every conjugacy class $g$ of hyperbolic matrices in $\GL_2(\Z)$ contains reduced
representatives, all of which have the same length $\ell(g)$. The number
of such representatives is $\ell(g)/k(g)$ where $k(g)$ is the largest integer such
that $g=h^{k(g)}$ for some $h$, \cite{LewZag}.

\smallskip

\begin{lem}\label{BratRMg}
Let $\cA_\theta$ be a quantum   torus with real multiplication, namely
$\theta$ is a quadratic irrationality. Then the AF algebra $\bA_\theta$ in which
$\cA_\theta$ embeds is determined by a stationary Bratteli diagram with $\# V_n=2$
for all $n\geq 1$ and with incidence matrix $\varphi_n=\varphi \in {\rm Red}\subset {\rm GL}_2(\Z)$
that fixes the modulus, 
$$ \varphi(\theta)=\frac{a\theta +b}{c\theta +d } = \theta, \ \  \ \text{ where }
\ \  \varphi= \left( \begin{array}{cc} a & b \\ c & d \end{array}\right). $$
\end{lem}

\proof Let $\bA_\theta$ be the AF algebra with $\bA_n = \cM_{q_n}(\C)\oplus \cM_{q_{n-1}}(\C)$,
as above, in which the quantum   torus $\cA_\theta$ embeds. The AF algebra 
$\bA_\theta$ is determined by a Bratteli diagram $\Gamma_\theta$ of the form
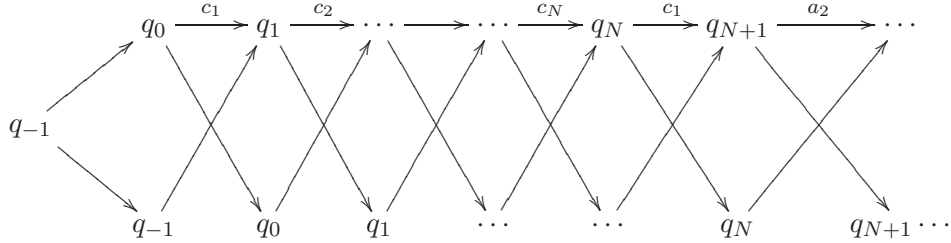
\[\xymatrix{
& q_0 \ar[r]^{c_1} \ar[rdd] & q_1 \ar[r]^{c_2} \ar[rdd] & \cdots \ar[r] \ar[rdd] & 
\cdots  \ar[r]^{c_N}\ar[rdd] & q_N \ar[r]^{c_1} 
\ar[rdd] & q_{N+1} \ar[r]^{a_2} \ar[rdd] & \cdots \\
q_{-1} \ar[ur] \ar[dr] \\
& q_{-1} \ar[ruu]  & q_0 \ar[ruu] & q_1 \ar[ruu] & \cdots\ar[ruu]   & \cdots \ar[ruu] & q_N \ar[ruu] & q_{N+1} \cdots}  \]
where $q_{-1}=q_0=1$ and $q_n = c_n q_{n-1} + q_{n-2}$ are the ranks of
the matrix algebras attached to the vertices.

\smallskip

If $\theta \in \R$ is a quadratic irrationality, then there exists a matrix $g \in {\rm SL}_2(\Z)$
such that $g\cdot \theta = \theta$ with $g$ acting by fractional linear transformations. 
Moreover, the continued
fraction expansion of $\theta$ is eventually periodic. Let $c_1, \ldots, c_N$
be the period of the continued fraction expansion of $\theta$. 

\smallskip

We telescope the Bratteli diagram $\Gamma_\theta$ to a new $\Gamma'_\theta$
that collapses together $N$ successive steps from $V_1$ to $V_N$, and then the
next  $N$ steps and so on, so that in $\Gamma'_\theta$ we have
$\varphi'_{\ell N+n} = \prod_{k=1}^N \varphi_k$, for all $\ell\geq 0$. 
Then the resulting diagram is isomorphic to the stationary diagram 
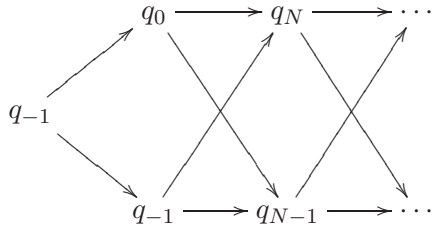
\[\xymatrix{
& q_0 \ar[r] \ar[rdd] & q_N \ar[r] \ar[rdd]  & \cdots \\
q_{-1} \ar[ur] \ar[dr] \\
& q_{-1} \ar[r] \ar[ruu]  & q_{N-1} \ar[r] \ar[ruu] & \cdots  }
\]
with incidence matrix
$$ g = \left( \begin{array}{cc} a & b \\ c & d \end{array}\right) \in {\rm Red}\subset {\rm GL}_2(\Z). $$
\endproof

\smallskip

\begin{ex}\label{exAFcontfr}{\rm 
In particular, in the case of with $\theta=\tau$,
the golden ratio, we find a copy of the quantum   torus $\cA_\tau$
embedded in the Fibonacci AF algebra.
The quantum   torus $\cA_\tau$ with modulus the golden ratio $\tau = (1+\sqrt{5})/2$
embeds in the Fibonacci AF $C^*$-algebra $\bA$ of Proposition \ref{propBrat}, inducing
an isomorphism on $K_0$. Indeed, 
the golden ratio has continued fraction expansion $\tau=[1,1,1,1,1,\ldots]$, hence
we recognize the Fibonacci AF algebra $\bA$ as the AF algebra of \cite{PiVo} 
for the quantum   torus $\cA_\tau$.}
\end{ex}

\medskip
\subsection{Bimodules and quantum tori with real multiplication}

As above, consider the action of ${\rm GL}_2(\Z)$ on $\P^1(\R)$ by fractional linear
transformations 
$$ g(\theta) = \frac{a\theta +b}{c \theta +d}, \ \  \ \text{ for }  \ \ \
g = \left( \begin{array}{cc} a & b \\ c & d \end{array}\right) \in {\rm GL}_2(\Z). $$

\smallskip

For $\theta \in \R\smallsetminus \Q$, an important family of finite
projective (right) modules over the quantum   torus $\cA_\theta$ 
was constructed in \cite{Co}. They are defined as follows: 
given a matrix
\begin{equation}\label{gabcd}
 g = \left( \begin{array}{cc} a & b \\ c & d \end{array}\right) \in {\rm GL}_2(\Z)  
\end{equation} 
one considers the Schwartz space
$\cS(\R\times \Z/c\Z)=\cS(\R)^c$ with an action of the generators $U$
and $V$ given by
$$ (fU)(t,s)=f(t- \frac{c\theta +d}{c}, s-1), \ \ \  (fV)(t,s)= \exp(2\pi i (t-\frac{ad}{c})) f(t,s). $$
This module is denoted by $E_g(\theta)$ and is referred to as a {\em basic module}.
It also carries a (left) action of $\cA_{g(\theta)}$, with generators $U'$ and $V'$, given by
$$ (U'f)(t,s)=f(t-\frac{1}{c}, s-a), \ \ \  (V'f)(t,s)=\exp(2\pi i( \frac{t}{c\theta+d} - \frac{s}{c})) f(t,s) . $$
Norm completion is taken with respect to $\| f \|=\| {}_{\cA_{g(\theta)}}\langle f,f \rangle \|^{1/2}$. 
The basic modules satisfy (Corollary 1.4 of \cite{PolSch})
\begin{equation}\label{HomEg}
{\rm Hom}_{\cA_\theta}(E_g(\theta),E_h(\theta))\simeq E_{hg^{-1}}(g(\theta)).
\end{equation}

\smallskip

In the real multiplication case, with $\theta$ a quadratic irrationality, we have a $g\in \GL_2(\Z)$
with $g(\theta)=\theta$. We can assume that $c\theta+d \geq 0$, for $g$ written as in \eqref{gabcd}.
The basic module $E_g(\theta)$ is an $\cA_\theta$--$\cA_\theta$ 
bimodule and one can form tensor products over $\cA_\theta$. The $n$-fold tensor
product satisfies
\begin{equation}\label{tensorbasic}
E_g(\theta)\otimes_{\cA_\theta} E_g(\theta) \otimes_{\cA_\theta} \cdots \otimes_{\cA_\theta}
E_g(\theta) \simeq E_{g^n}(\theta). 
\end{equation}
The bimodule $E_g(\theta)$ generates the nontrivial self
Morita equivalences of the torus $\cA_\theta$ that determine the real multiplication
structure, \cite{Manin}.

\smallskip

\begin{thm}\label{RManyons}
Let $\theta$ be a quadratic irrationality and $g\in \GL_2(\Z)$
a matrix with $g(\theta)=\theta$. Assume that $g$ has non-negative 
entries and $\det(g)=-1$. Let $\cA_\theta$ be the quantum 
torus with real multiplication implemented by the basic bimodule $E_g(\theta)$.
Then the objects $\cA_\theta$ and $E_g(\theta)$ in the category of
finite projective (right) modules over $\cA_\theta$ form an anyon system,
whose fusion ring is given by $\Z[x_0,x_1]$ modulo $x_i x_j=\sum_k N^k_{ij} x_k$,
with fusion matrices $N_{0j}^k=\delta_{jk}=N_{j0}^k$ and $N_{11}^1 ={\rm Tr}(g)$
and $N_{11}^0=1$. The elements $x_0, x_1$ in the fusion ring correspond
to the classes $x_0=[\cA_\theta]$ and $x_1=[E_g(\theta)]$ in $K_0(\cA_\theta)$.
\end{thm}

\proof
Consider the subcategory $\cM_\theta$ of the category of finite projective (right) modules 
over $\cA_\theta$ generated by the basic module $E_g(\theta)$ and by 
$\cA_\theta$, seen as a module over itself. 
This category $\cM_\theta$ is monoidal, with tensor product $\otimes_{\cA_\theta}$,
with the object $\cA_\theta={\bf 1}$ as the unit. 
To see that this gives a categorification of the fusion ring 
it suffices to check what the tensor product 
$E_g(\theta)\otimes_{\cA_\theta} E_g(\theta)=E_{g^2}(\theta)$ corresponds to
in $K_0(\cA_\theta)$. 
We have $K_0(\cA_\theta)=K_0(\bA_\theta)=\Z + \Z\theta$. The basic module
$E_g(\theta)$ has dimension (given by the range of the trace on $K_0$)
$c\theta +d$, \cite{Co}, hence under the identification $K_0(\cA_\theta)=\Z + \Z\theta$,
given by the von Neumann trace of the quantum torus, we have 
$[E_g(\theta)]= c\theta +d$. The unit element satisfies $[\cA_\theta]=[E_1(\theta)]=1$. 
We then have $[E_{g^2}(\theta)]=c'\theta + d'$, where
$$ g^2=\left( \begin{array}{cc} a' & b' \\ c' & d' \end{array}\right)=
\left( \begin{array}{cc}  a^2+bc & b(a+d)   \\  c(a+d)   & d^2 +bc 
\end{array}\right) $$
so that we get
$$ [E_{g^2}(\theta)]= c(a+d) \theta + d^2 +bc = 
{\rm Tr}(g) \, (c\theta+d) -\det(g) = {\rm Tr}(g) \,[E_g(\theta)]-\det(g)[\cA_\theta]. $$
If all the entries of $g$ are non-negative and $\det(g)=-1$, we obtain
$$ [E_{g^2}(\theta)]= {\rm Tr}(g) \,[E_g(\theta)] + [{\bf 1}]. $$
Thus, we can form an anyon system with two anyon types $x_0$and $x_1$,
where $X_0={\bf 1}$ and $X_1=E_g(\theta)$, with the tensor product \eqref{tensorbasic}.
The corresponding classes $x_0=[X_0]$ and $x_1=[X_1]$ in $K_0(\cA_\theta)$ 
generate a fusion ring with $x_0 x_i =x_i x_0 =x_i$ for $i=0,1$, and 
$x_1 x_1 ={\rm Tr}(g)\, x_1 + x_0$. 
\endproof

\smallskip

\begin{cor}\label{Smatrix}
Let $\theta$ be a quadratic irrationality and 
$$ g=\left(\begin{array}{cc} a & b \\ c & d \end{array}\right) \in \GL_2(\Z) $$ 
a matrix satisfying $g(\theta)=\theta$ with non-negative entries and $\det(g)=-1$.
Then for the anyon system constructed in Theorem \ref{RManyons}
the matrix $\tilde S$ of the modular fusion rules \eqref{NSi} is
of the form 
\begin{equation}\label{tildeSg}
 \tilde S =\frac{1}{(1+(c\theta +d)^2)^{1/2}} \left( \begin{array}{cc} 1 &  c\theta+d  \\
c\theta+d & -1 \end{array}\right).
\end{equation}
\end{cor}

\proof We write the fusion matrix in the form 
\begin{equation}\label{N1matTrg}
 N_1= \left( \begin{array}{cc} {\rm Tr}(g) & 1 \\ 1 &  0  \end{array}\right). 
\end{equation} 
The eigenvalues of $N_1$ are the same as the eigenvalues of $g$,
since they are solutions of the characteristic polynomials
$$ \det(g - \lambda\,I) = \lambda^2 - {\rm Tr}(g)\lambda + \det(g)= \det(N_1 - \lambda\, I) =0. $$
The eigenvalues satisfy the relation 
$\lambda_1 \lambda_2=\det(g)=-1$, and 
the eigenvectors of $N_1$ satisfy the conditions $y =\lambda x$ and
$x(1 + {\rm Tr}(g) \lambda - \lambda^2)=0$, hence we find that a unitary matrix of 
eigenvectors is of the form 
$$  \tilde S =\frac{1}{(1+\lambda^2)^{1/2}} \left( \begin{array}{lr} 1 &  \lambda  \\
\lambda & -1 \end{array}\right), $$
with $\lambda$ an eigenvalue of the matrix $g$.
Observe then that 
$$ g \left(\begin{array}{c} \theta \\ 1 \end{array}\right) = 
\left( \begin{array}{c} a\theta + b \\ c\theta+ d \end{array}\right)=
(c\theta + d)\, \left( \begin{array}{c} \theta \\ 1 \end{array}\right), $$
since $g(\theta)=\theta$, hence $\lambda= c\theta+d$ is an eigenvalue of $g$. Thus,
we obtain \eqref{tildeSg}.
\endproof

\smallskip

In a similar way, we can show that the AF algebras associated to the incidence
matrices $g$ and \eqref{N1matTrg} are the same.

\smallskip

\begin{cor}\label{Trgincidence}
Let $\theta$ be a quadratic irrationality with $\theta>1$.
Let $\bA_\theta$ be the AF algebra computed by the stationary Bratteli diagram with 
$\#V_n=2$ for all $n\geq 1$ and incidence matrix $\varphi_n=\varphi=g$ with 
$g(\theta)=\theta$, where $g$ has non-negative entries and $\det(g)=-1$. Then $\bA_\theta$
 is isomorphic to the AF algebra computed by the stationary
Bratteli diagram with incidence matrix
\begin{equation}\label{phiTr}
\varphi= \left(\begin{array}{cc} {\rm Tr}(g) & 1 \\ 1 & 0  \end{array}\right).
\end{equation}
\end{cor}

\proof It suffices to show that the ordered $K_0$-groups of the two AF algebras are
isomorphic. They're both isomorphic to $\Z^2$ as abelian groups, so we need to
check that the order structure agrees. The order structure on the direct limit of
$$ \Z^2 \stackrel{g}{\to} \Z^2 \stackrel{g}{\to} \Z^2 \stackrel{g}{\to} \Z^2 \to \cdots $$
is obtained by describing the direct limit as $K_0(\bA_\theta)=\cup_m G_m$ with 
$G_m=g^{-1}(\Z^2)$ with $G_m^+=g^{-1}(\Z^2_+)$. An element $h$ is in the positive
cone $K_0(\bA_\theta)_+$ if it is in some $G_m^+$. Equivalently, the trace of $h$ is
positive, where the trace on $G_m$ that induces the trace on $K_0(\bA_\theta)$ 
is given in terms of Perron--Frobenius eigenvalue and eigenvector of $g$,
(see (11)--(13) of \cite{BJKR}) as $\lambda^{-m+1}\langle v, h\rangle$. 
As observed in Corollary \ref{Smatrix} above,
$g$ has a positive eigenvalue $\lambda=c\theta+d$ with eigenvector $(\theta,1)$.
Thus, we have $h=(n,m)$ positive if $\theta n +m \geq 0$. In the case of the
stationary Bratteli diagram 
$$ \Z^2 \stackrel{N_1}{\to} \Z^2 \stackrel{N_1}{\to} \Z^2 \stackrel{N_1}{\to} \Z^2 \to \cdots $$
with $N_1$ as in \eqref{N1matTrg}, we similary have the condition that $h=(n,m)$ is in the 
positive cone determined by the positivity of $\langle v, h\rangle$. The matrix $N_1$
has the same eigenvalues as $g$, hence the same positive eigenvalue $\lambda=c\theta+d$,
but with the corresponding eigenvector given by $(1,c\theta+d)$ as we saw in 
Corollary \ref{Smatrix} above. Thus, $h=(n,m)$ is in the positive cone whenever 
$n+ m(c\theta+d)\geq 0$. To see that these two conditions are equivalent, if 
$n+ m(c\theta+d)\geq 0$ then $m+ n/(c\theta+d)\geq 0$ and $n\leq a\theta n + bn$ since
$\theta>1$ and $a,b\geq 0$ (not both zero), hence using $g(\theta)=\theta$ we obtain
$n\theta +m \geq 0$. Conversely, if $n\theta +m \geq 0$, then $n \theta + (a\theta+b)m\geq 0$,
hence $m+n/(c\theta+d)\geq 0$ which gives that $h$ is in the positive cone $n + m(c\theta+d)\geq 0$. 
\endproof

\smallskip

Corollary \ref{Trgincidence} ensures that the Hilbert space determined by the Bratteli diagram
of the AF algebra $\bA_\theta$ is indeed isomorphic to the Hilbert space 
of the fusion paths of the anyon system constructed in 
Theorem \ref{RManyons} above.

\smallskip

\begin{cor}\label{FiboTrg}
Theorem \ref{RManyons} applied to the case of the golden ratio $\theta=\tau=(1+\sqrt{5})/2$
recovers the Fibonacci anyon system.
\end{cor}

\proof In this case we have $g(\tau)=\tau$ with
$$ g = \left(\begin{array}{cc} 1 & 1 \\ 1 & 0 \end{array}\right) $$
with ${\rm Tr}(g)=1$ and $\det(g)=-1$. The basic bimodule $E_g(\tau)$
satisfies $[E_{g^2}(\tau)]=[E_g(\tau)\otimes_{\cA_\theta} E_g(\tau)]=
[E_g(\tau)]+[{\bf 1}]$. Thus, the fusion ring generated by $x_0=[\cA_\tau]=[{\bf 1}]$
and $x_1=[E_g(\tau)]$ satisfies the fusion rules of the Fibonacci anyons
$x_0 x_i = x_i x_0 = x_i$ and $x_1 x_1 = x_1 + x_0$. This ring structure
agrees with the product on $K_0(\cA_\tau)=\Z+\Z\tau$ seen as algebraic
integers in $\bK=\Q(\sqrt{5})$, with $[E_g(\tau)]=\tau$ and $[{\bf 1}]=1$ and
$[E_{g^2}(\tau)]=\tau+1 =\tau^2$.
\endproof

Thus, we can view the construction above for more general quadratic
irrationalities $\theta$ as a direct generalization of the Fibonacci anyons.

\smallskip

\begin{rem}\label{noMTC}{\rm 
By Corollary \ref{Smatrix} and the classification of the modular tensor categories 
of \S 3 of \cite{RoStoWa},
we see that the modular fusion rules of the real multiplication anyons of Theorem \ref{RManyons}
arise from a modular tensor category only in the case of the Fibonacci anyons, with
$\theta=\tau=(1+\sqrt{5})/2$. In all the other cases one has modular fusion rules, but the
corresponding categorification has weaker properties than the modular tensor case.}
\end{rem}

\smallskip
\subsection{Quantum gates and approximate generators of the quantum   torus}\label{QGatesSec}

One can realize the generators of a quantum   torus $\cA_\theta$, through the 
embedding into the AF algebra $\bA_\theta$, as
limits of a sequence of matrices, which give approximate generators. We show
that the natural choice of an approximating sequence (see \cite{LaLiSza}, \cite{PiVo})
can be interpreted as quantum gates.

\smallskip

\begin{prop}\label{qgatesUVn}
Let $\theta$ be a quadratic irrationality and $\bA_\theta$ the AF algebra
constructed from its continued fraction expansion as above. Then $\bA_\theta$
contains unitary operators $U_n$ and $V_n$ that approximate the generators
$U$ and $V$ of the quantum torus $\cA_\theta$ and that act on the Hilbert
space $\C^{q_n}$ of fusion paths of length $n$ as phase shifter gates with phase 
$\exp(2\pi i \frac{p_n}{q_n})$ and downshift permutation gates on $q_n$ elements,
respectively.
\end{prop}

\proof
As shown in \cite{PiVo},
the generators $U$ and $V$ of the quantum   torus are approximated by
elements $U_n$ and $V_n$ in $\cM_{q_n}(\C)$ satisfying the relation
\begin{equation}\label{UVn}
U_n\, V_n = \exp(2\pi i \frac{p_n}{q_n}) \, V_n\, U_n,
\end{equation}
with $p_n/q_n$ the successive quotients of the continued fraction expansion
approximation of the modulus of the quantum   torus. The elements $U_n$
and $V_n$ are explicitly given by $q_n\times q_n$-matrices
\begin{equation}\label{Un}
U_n =\left(\begin{array}{ccccc} 1 & & & & \\
& \xi_n & & & \\
& & \xi_n^2 & & \\
 & & & \ddots & \\
& & & & \xi_n^{q_n-1}
\end{array}\right), \ \ \ \text{ with } \ \  \ \xi_n =\exp(2\pi i \frac{p_n}{q_n}),
\end{equation}
\begin{equation}\label{Vn}
V_n =\left(\begin{array}{cccccc} 
0 & 1 & 0 & \cdots & 0 & 0 \\
0 & 0 & 1 & \cdots & 0 & 0 \\
\vdots & & & & & \vdots \\
0 & 0 & 0 & \cdots & 0 & 1 \\
1 & 0 & 0 & \cdots & 0 & 0
\end{array}\right)
\end{equation}

\smallskip

Thus, the approximate generators $U_n$ and 
$V_n$ act on the Hilbert space $\C^{q_n}$ of the system
as (compositions of) phase shifter gates with phase $\xi_n$ and downshift 
permutation gates on $q_n$ elements (see e.g.~\S 3.3.3 of \cite{Will}).
\endproof

\medskip
\subsection{$F$-matrices and pentagons}
We consider the problem of computing the $F$-matrix for our
real multiplication anyon systems. 

\smallskip

Recall that the $F$-matrices
are determined by the natural transformations 
\begin{equation}\label{HomFmat}
\bigoplus_c \Hom(x_u, x_c \otimes x_k)\otimes \Hom(x_c, x_i \otimes x_j) 
\stackrel{F^{ijk}_u}{\longrightarrow}
\bigoplus_d \Hom(x_u, x_i \otimes x_d) \otimes \Hom(x_d, x_j \otimes x_k),
\end{equation}
where the left-hand-side is $\Hom(x_u, (x_i \otimes x_j)\otimes x_k)$ and the
right-hand-side is $\Hom(x_u, x_i \otimes (x_j \otimes x_k))$.

\medskip
\subsubsection{$F$-matrices and basic modules}

In our setting, we can view the transformation \eqref{HomFmat} at different levels. 
If we view as in Theorem \ref{RManyons} the fusion of
anyons as the tensor product over $\cA_\theta$ of basic modules,
we obtain the following description of the $F$-matrices.

\smallskip

\begin{prop}\label{FmatEgtheta}
Let $\theta$ be a quadratic irrationality and  $g\in \GL_2(\Z)$ 
with nonnegative entries $\det(g)=-1$, such that $g(\theta)=\theta$.
Let $X_0=\cA_\theta$ and $X_1=E_g(\theta)$, and set
\begin{equation}\label{HomFmatEg}
\bigoplus_c \Hom(X_u, X_c \otimes X_k)\otimes \Hom(X_c, X_i \otimes X_j) 
\stackrel{F^{ijk}_u}{\longrightarrow}
\bigoplus_d \Hom(X_u, X_i \otimes X_d) \otimes \Hom(X_d, X_j \otimes X_k),
\end{equation}
with $\Hom=\Hom_{\cA_\theta}$ and $\otimes=\otimes_{\cA_\theta}$.  Then
\begin{equation}\label{FijkuEg}
F^{ijk}_u:  E_h(\theta) \oplus E_h(\theta) \to E_h(\theta) \oplus E_h(\theta),
\end{equation}
for $h=g^k$, where $k$ is the difference between the number of upper indices
equal to $1$ and the number of lower indices equal to $1$ in $F^{ijk}_u$.
\end{prop}

\proof All indices $i,j,k,u$ take value either zero or one. All the sixteen resulting
cases are checked similarly, using the fact that
$$  {\rm Hom}_{\cA_\theta}(E_g(\theta),E_h(\theta)\simeq E_{h g^{-1}}(g(\theta)). $$
For instance, for $F^{101}_0$ we have as source
$$ \Hom(X_0,X_0\otimes X_1)\otimes \Hom(X_0,X_1\otimes X_0)
\oplus \Hom(X_0,X_1\otimes X_1) \otimes \Hom(X_1,X_1\otimes X_0) = $$
$$ E_g(\theta) \otimes_{\cA_\theta} E_g(\theta) \oplus E_{g^2}(\theta) \otimes_{\cA_\theta} \cA_\theta
= E_{g^2}(\theta)  \oplus E_{g^2}(\theta) $$
and as target
$$ \Hom(X_0,X_1\otimes X_0)\otimes \Hom(X_0,X_0\otimes X_1) \oplus
\Hom(X_0, X_1\otimes X_1) \otimes \Hom(X_1, X_1\otimes X_0) = $$
$$ E_g(\theta) \otimes_{\cA_\theta} E_g(\theta) \oplus E_{g^2}(\theta) \otimes_{\cA_\theta} \cA_\theta
= E_{g^2}(\theta)  \oplus E_{g^2}(\theta). $$
The other cases are checked similarly.
\endproof

\smallskip

We know by construction that, if $g\in \GL_2(\Z)$ is a matrix
with nonnegative entries and with $\det(g)=-1$ satisfying
$g(\theta)=\theta$, then the modules $E_{g^2}(\theta)$ and 
$E_g(\theta)^{\oplus {\rm Tr}(g)}\oplus \cA_\theta$ both have
the same class $[E_{g^2}(\theta)]={\rm Tr}(g)\, [E_g(\theta)]+ 1$ 
in $K_0(\cA_\theta)$. Thus, we can define $F$-matrices in
a different way, by formally replacing the basic module 
$E_{g^2}(\theta)$ with $E_g(\theta)^{\oplus {\rm Tr}(g)}\oplus \cA_\theta$.
This leads to a definition of the $F$-matrices that is more similar
to the usual setting for anyon systems, where the $F$-matrix 
$F^{ijk}_u$ is an $m\times m$-matrix for 
$m = N^u_{0k} N^0_{ij} + N^u_{1k} N^1_{ij} = N^u_{i0} N^0_{jk} + N^u_{i1} N^1_{jk}$.

\smallskip

\begin{prop}\label{FETrg}
Let $\theta$ be a quadratic irrationality and  $g\in \GL_2(\Z)$ 
with nonnegative entries $\det(g)=-1$, such that $g(\theta)=\theta$.
Let $X_0=\cA_\theta$ and $X_1=E_g(\theta)$, but with the modified fusion rule
$X_1\otimes X_1=E_g(\theta)^{\oplus {\rm Tr}(g)}\oplus \cA_\theta
=X_1^{\oplus {\rm Tr}(g)}\oplus X_0$. Then the $F$-matrices are as
in Proposition \ref{FmatEgtheta}, except for the cases $F^{110}_0$, $F^{101}_0$,
$F^{011}_0$, $F^{111}_1$, $F^{111}_0$, and $F^{0 0 0}_1$.
The cases $F^{110}_0$, $F^{101}_0$,
$F^{011}_0$, and $F^{111}_1$ are endomorphisms of
$E_g(\theta)^{\oplus 2 {\rm Tr}(g)} \oplus \cA_\theta^{\oplus 2}$, and the case
$F^{111}_0$, which is an endomorphism of
$E_g(\theta)^{\oplus 2({\rm Tr}(g^2)+1)} \oplus \cA_\theta^{\oplus 2{\rm Tr}(g)}$.
However, the remaining case $F^{111}_0$ is in general not compatible
with this formulation, though in the Fibonacci case it can
be interpreted as an endomorphism of a finite projective module with
trace $\tau=(1+\sqrt{5})/2$.
\end{prop}

\proof In the cases $F^{110}_0$, $F^{101}_0$,
$F^{011}_0$, and $F^{111}_1$ we have a term $X_1\otimes X_1$ that
occurs on both sides of \eqref{HomFmatEg}, so that both sides 
are given  by the direct sum of two copies of 
$E_g(\theta)^{\oplus {\rm Tr}(g)}\oplus \cA_\theta$. In the case of 
$F^{111}_0$, which in the setting of Proposition \ref{FmatEgtheta} 
has a term $E_{g^3}(\theta)$ on both sides, which in this case
gets replaced by a direct sum of copies of $E_g(\theta)$ and $\cA_\theta$
with the same class in $K_0(\cA_\theta)$. The class is given by
$[E_{g^3}(\theta)]=c'\theta +d'$, where 
$$ g^3 = \left(\begin{array}{cc} a' & b' \\ c' & d' \end{array}\right) =
\left(\begin{array}{cc} a^3+abc+bc(a+d) & a^2b +b^2c + bd(a+d)\\
ac(a+d)+d^2 c+ b c^2 & bc (a+d) + d^3 + bcd \end{array}\right). $$
Thus we have
$$ [E_{g^3}(\theta)]= (a^2 +d^2 +ad + bc) (c\theta+d) + abc +dbc -da^2 -ad^2 =
({\rm Tr}(g^2) -\det(g)) (c\theta +d) - {\rm Tr}(g) \det(g). $$
Under the assumption that $\det(g)=-1$ this gives that
$$ [E_{g^3}(\theta)]=[ E_g(\theta)^{\oplus ({\rm Tr}(g^2)+1)} \oplus \cA_\theta^{\oplus {\rm Tr}(g)} ]. $$
In the case of $F^{111}_0$ we have a sum of two copies of $E_{g^{-1}}(\theta)$ on both
sides of \eqref{HomFmatEg}. The $K_0$-class is $[E_{g^{-1}}(\theta)]=c\theta -a=
[E_g(\theta)] -{\rm Tr}(g)$, which is not the class of a direct sum $E_g(\theta)^{\oplus a}\oplus
\cA_\theta^{\oplus b}$ for any $a,b\in \Z_{\geq 0}$. In the Fibonacci case one has 
$[E_g(\tau)] =\tau+1$ and $[E_{g^{-1}}(\tau)]=\tau$.
\endproof

\medskip
\subsubsection{Pentagon relations}

The pentagon relations between the $F$-matrices arise from rearranging the fusion order
in a fusion tree of five anyons. The relations can be written in the form
\begin{equation}\label{Pentagon}
(F^{i_a j_3 k_4}_{u_5})^c_b \, (F^{i_1 j_2 k_c}_{u_5})^d_a =
\sum_e (F^{i_1 j_2 k_3}_{u_b})^e_a\, (F^{i_1 j_e k_4}_{u_5})^d_b \,
(F^{i_2 j_3 k_4}_{u_d})^c_e,
\end{equation}
where in a term of the form $(F^{ijk}_u)^b_a$ the labels $a$ and $b$ denote the
internal edges of the trees that are exchanged in the rearranging of the fusion tree
and the labels $\{ i,j,k, u \}$ are the anyons types (values $0$ or $1$) assigned to
the three inputs and one output of the edge labelled $a$ in the first tree.

\smallskip

Similarly, there is a hexagon relation involving the $F$-matrix determined
by the pentagon relation \eqref{Pentagon} and the braiding $R$-matrix,
with
\begin{equation}\label{Hexagon}
\sum_b (F^{i_2 j_3 k_1}_{u_4})^d_a\, R_{u_4}^{i_1 j_b}\, 
(F^{i_1 j_2 k_3}_{u_4})^b_a = R_{u_c}^{i_1 j_3} \, (F_{u_4}^{i_2 j_1 k_3})^c_a R_{u_a}^{i_1 j_2}.
\end{equation}

\medskip

\begin{ex}\label{FibF} {\rm
In the usual setting of Fibonacci anyons, one interprets the $F$-matrices and the
pentagon relations as equations for unitary matrices acting on a finite dimensional
Hilbert space. The only two matrices that are possibly nontrivial in the Fibonacci case 
are $F^{ 1 1 1}_0=t$ and 
$$ F^{1 1 1}_1 = \left(\begin{array}{cc} p & q \\ r & s \end{array}\right), $$
respectively of rank $N^0_{01}N^0_{11}+N^0_{11}N^1_{11}= 1$ and
$N^1_{01}N^0_{11}+N^1_{11}N^1_{11}=2$, and the pentagon equations are then given by
(\cite{Wang}, Example 6.4)
$$ \left(\begin{array}{cc} 1 & 0 \\ 0 & t \end{array}\right) 
\left(\begin{array}{cc} 1 & 0 \\ 0 & t \end{array}\right) =
\left(\begin{array}{cc} p & q \\ r & s \end{array}\right) \left(\begin{array}{cc} 1 & 0 \\ 0 & t \end{array}\right)
\left(\begin{array}{cc} p & q \\ r & s \end{array}\right)  $$
$$ \left(\begin{array}{ccc} 1 & 0 & 0 \\ 0 & p & q \\ 0 & r & s \end{array}\right) 
\left(\begin{array}{ccc} 0 & 1 & 0 \\ 1 &  0 & 0 \\ 0 & 0 & 1 \end{array}\right)
\left(\begin{array}{ccc} 1 & 0 & 0 \\ 0 & p & q \\ 0 & r & s \end{array}\right) = 
 \left(\begin{array}{ccc} p & 0 & q \\ 0 & t & 0 \\ r & 0 & s \end{array}\right)
\left(\begin{array}{ccc}  1 & 0 & 0 \\ 0 & p & q \\ 0 & r & s \end{array}\right) 
\left(\begin{array}{ccc}  p & 0 & q \\ 0 & t & 0 \\ r & 0 & s \end{array}\right), $$
which, together with the unitarity constraint, have solution $F^{ 1 1 1}_0=1$ and
$$ F^{1 1 1}_1 = \left(\begin{array}{cc} \tau^{-1} & \tau^{-1/2} \\ \tau^{-1/2} & -\tau^{-1} 
\end{array}\right), $$
with $\tau=(1+\sqrt{5})/2$ the golden ratio. The $R$-matrix is then determined by
the $F$-matrix and by \eqref{Hexagon} and it is given by
$$ R = \left(\begin{array}{cc} \exp(\frac{4\pi i}{5}) & 0 \\
0 & - \exp(\frac{2\pi i}{5})  \end{array}\right). $$}
\end{ex}

\smallskip

More generally, one expects matrix equations for matrices
$F^{ijk}_u$ of size $N^u_{0k} N^0_{ij} + N^u_{1k} N^1_{ij}$, with
$N^k_{0j} =N^k_{j0}= \delta_{jk}$ and
$$ N^k_{1j}=\left( \begin{array}{cc} {\rm Tr}(g) & 1 \\ 1 & 0 \end{array} \right). $$

\medskip
\subsection{Pentagon relations in the quantum torus and quantum dilogarithm}

When we interpret the $F$-matrices as homomorphisms between sums of
basic modules as in \eqref{FijkuEg}, the matrix elements $(F^{ijk}_u)^b_a$
are homomorphims 
$$ (F^{ijk}_u)^b_a \in {\rm Hom}_{\cA_\theta}(E_h(\theta), E_h(\theta))\simeq \cA_\theta, $$
hence we regard \eqref{Pentagon} as an equation in $\cA_\theta$. 

\smallskip

This means that, in principle, solutions to \eqref{Pentagon} in $\cA_\theta$
can be constructed from elements of the algebra that are known to satisfy 
other types of pentagon relations. We make a proposal here for a family
of approximate solutions, related to the Faddeev--Kashaev quantum
dilogarithm \cite{FadKas}, through the approximation of the generators of
the quantum torus $\cA_\theta$ by elements of the AF algebra $\bA_\theta$
described in \S \ref{QGatesSec}.

\smallskip

The quantum dilogarithm function was introduced in \cite{FadKas}
 (see also \cite{BaRe}) as a function that provides a quantized version of the 
Rogers pentagon identity for the dilogarithm function, to which it reduces
in the limit of the quantization parameter $q\to 1$. The quantum dilogarithm
is originally defined as an element in a completion of the Weyl algebra
generated by invertibles $U$ and $V$ with the relation $UV = q VU$, 
for a fixed $q\in \C^*$ with $|q|<1$, as the function
\begin{equation}\label{QDilog}
\Psi_q(x) = (x;q)_\infty =\prod_{k=0}^\infty (1-q^k x).
\end{equation} 
It is shown in \cite{FadKas} that it satisfies the pentagon identity
\begin{equation}\label{QDilogPenta}
\Psi_q(V) \, \Psi_q(U) = \Psi_q(U)\, \Psi_q(-VU)\, \Psi_q(V).
\end{equation}
The infinite product \eqref{QDilog} is no longer convergent when the parameter
$q$ is on the unit circle $|q|=1$. However, there is a way to extend the quantum
dilogarithm to the case where $q$ is a root of unity, in such a way that it still
satisfies a pentagon relation, \cite{BaRe}, \cite{FadKas}. 
For $q=\exp(2\pi i p_n/q_n)=\xi_n$, we can consider the completion of the Weyl algebra
as being the rational quantum torus, with generators
$U = u U_n$ and $V= v V_n$, with $U_n$ and $V_n$ the $q_n\times q_n$
matrices as in  \S \ref{QGatesSec}, and with $u,v \in S^1$. 

For $\zeta$ a root of unity of order $N$, the quantum dilogarithm
is then defined using the function $\Phi_\zeta(x)$ of the form
\begin{equation}\label{QDilogzeta}
\Phi_\zeta(x) =(1-x^N)^{(N-1)/2N} \prod_{k=1}^{N-1} (1-\zeta^k x)^{-k/N}.
\end{equation}
and for $\zeta=\xi_n=\exp(2\pi i p_n/q_n)$ one obtains the pentagon relation (see
(3.18) of \cite{FadKas})
\begin{equation}\label{QDilogzetapenta}
\begin{array}{l}\displaystyle{
\Phi_{\xi_n}(v V_n) \, \Phi_{\xi_n}(u U_n) = } \\
\displaystyle{
\Phi_{\xi_n}\left(\frac{u}{(1-v^{q_n})^{1/q_n}} U_n\right)\,
\Phi_{\xi_n}\left(\frac{-uv}{(1-u^{q_n}-v^{q_n})^{1/q_n}} U_n V_n\right) \, 
\Phi_{\xi_n}\left(\frac{v}{(1-u^{q_n})^{1/q_n}} V_n\right).}
\end{array}
\end{equation}
For a fixed choice of $u,v$, with $u^{q_n}\neq 1$ and $v^{q_n}\neq 1$, we can 
regard these as elements of the matrix algebra $\cM_{q_n}(\C)$, expressed in
terms of the approximate generators of the quantum torus $\cA_\theta$
embedded in the AF algebra $\bA_\theta$. As elements of $\bA_\theta$ they
can also be seen as operators acting on the Hilbert space of anyon fusion paths.

Thus, we propose to look for solutions of the equation \eqref{Pentagon} in $\cA_\theta$,
by setting the coefficients to be either zero or functions $\Phi_{\xi_n}(x)$ and
construct approximate solutions given by elements in $\bA_\theta$ that are
functions of the approximate generators $U_n$ and $V_n$ of the quantum
torus, satisfying pentagon relations of the form \eqref{QDilogzetapenta}. 

\bigskip
\subsection{Dimension functions and braiding}

In the original case of the Fibonacci anyons, where we have $F$ and $R$-matrices
as in Example \ref{FibF}, there are unitary operators 
$B_{i,n}$ describing the transformation the braid group element $\sigma_i$ 
affects on the physical Hilbert space for an $n$ anyon system, known as the ``braid matrix"; see \cite{NSSFS}, \cite{Pres}. If $i>n-1$, let $B_{i,n}=I$. Note that $\{\sigma_i, |\, i\in \mathbb{N}_0 \}$ generate $B_\infty$. It is shown in \cite{NSSFS}, \cite{Pres} that one can always determine $B_{i,n}$ for any $i,n \in \mathbb{N}_0$ in terms of ``rotation" and ``fusion" matrices, through the matrices $R$ and $F^{-1}RF$. We now define the action of $\sigma_i$ on some irreducible, $Fib(n+1)$-dimensional element $M$ of the AF algebra $\bA_\tau$ as $\sigma_i \cdot M = B_{i,n}MB_{i,n}^{-1}$. The dimension of all finite elements in $\bA_\tau$ is of the form $Fib(m), m\in \mathbb{N}_0$, and an $n$-anyon system corresponds to a $Fib(n+1)$ dimensional Hilbert space. We then define the action of a braid group element $\sigma_i$ on a reducible element 
$N$ of $\bA_\tau$ to be the direct sum of the actions of $\sigma_i$ on the irreducible parts of $N$. 
One can check that this action satisfies the aforementioned properties.

\medskip

In the case of the Fibonacci anyons, 
we can then show how one can associate disconnected 
braidings of the anyon system to elements of $K_0(\cA_\tau)=K_0(\bA_\tau)$, 
by constructing a representation of the $K_0$-group in the 
infinite braid group $B_\infty$.

\smallskip

As we have seen (see \cite{Dav}, \cite{PiVo2}), for $\cA_\tau$ the $K_0$ group, with
its positive cone, is given by
\begin{equation}\label{K0Atau}
(K_0(\cA_\tau),K_0^+(\cA_\tau))=
(\mathbb{Z}^2, \{ (n,m)\in \mathbb{Z}^2 \, |\, n+\tau m\geq 0\}).
\end{equation}
The range of the trace on $K_0(\cA_\tau)$ is given by the subgroup (pseudolattice)
$\mathbb{Z}+\tau \mathbb{Z}\subset \R$. 
We use here a description of the ordered $K_0$-group in terms of dimension functions.

\smallskip

\begin{defn}\label{defdimfunc}
A dimension function on a Bratteli diagram $\Gamma$ as defined in \cite{Kerov} 
is any $\mathbb{Z}$-valued function which is defined for almost all vertices 
of the graph and satisfies the equation
\begin{equation}\label{dimfunct}
 f(\Lambda) = \sum_{\lambda :\, \lambda \nearrow \Lambda}f(\lambda )\varkappa(\lambda , \Lambda),
 \ \ \ \Lambda \in \Gamma 
\end{equation} 
where $\varkappa$ is the multiplicity of the edge from $\lambda$ to $\Lambda$. 
Two functions $f_1, f_2$ are identified if they differ only on a finite set of vertices. A function is virtually positive if it is nonnegative on almost all vertices. 
\end{defn}

\smallskip

\begin{prop}\label{dimK0Binfty}
Let $\Gamma$ be the Bratteli diagram describing the Fibonacci anyons system, 
as in Proposition \ref{propBrat}. The dimension functions on $\Gamma$ are of the form
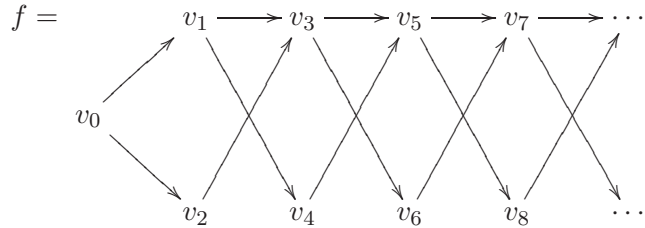
\[f =\xymatrix{
& v_1 \ar[r] \ar[rdd] & v_3 \ar[r] \ar[rdd] & v_5 \ar[r] \ar[rdd] & v_7 \ar[r] \ar[rdd] & \cdots \\
v_0 \ar[ur] \ar[dr] \\
& v_2 \ar[ruu]  & v_4 \ar[ruu] & v_6 \ar[ruu] & v_8 \ar[ruu] & \cdots} \]
where the $v_{i} \in \mathbb{Z}$ satisfy $v_{2k+1}=v_{2k}+v_{2k-1}$
and $v_{2k+2}=v_{2k-1}$ for $k>0$ and $v_2=v_0$. Let $B_\infty$ denote the
infinite braid group. Setting
\[ \varphi\, :\, \textrm{dimension functions on $\Gamma$} \rightarrow B_\infty \]
\begin{equation}\label{phifsigma}
 \varphi(f)=\sigma_1^{v_0}\sigma_3^{v_1}\sigma_5^{v_2}\cdots \sigma_{2n+1}^{v_n}\cdots 
\end{equation} 
determines a representation of $K_0(\cA_\tau)$ in $B_\infty$.
\end{prop}

\proof
It is proved in \cite{Kerov} that, for an AF algebra $\bA$ determined by a Bratteli diagram 
$\Gamma$, the group $K_0(\bA)$ is isomorphic to the group of dimension functions on 
$\Gamma$, and the cone $K_0^+(\bA)$ is isomorphic to the group of virtually positive elements.

\smallskip

In the case of the Fibonacci AF algebra, since the edges in $\Gamma$ all have multiplicity one, 
dimension functions satisfy
\[ f(\Lambda) = \sum_{\lambda :\, \lambda \nearrow \Lambda}f(\lambda ),\ \ \  \Lambda \in \Gamma, \]
hence they are given by functions as in the statement. 

\smallskip

Note that, given the values at any two vertices of the dimension function, we can determine 
all of the values of the dimension function up to some finite number of vertices (some of which 
may need to remain undefined due to the requirement that $v_i\in \mathbb{Z}$). Given some 
dimension function $f$, let $\lceil f \rceil$ denote the dimension function equivalent to $f$ that 
is defined on the maximal number of vertices, which can be uniquely constructed from $f$ via extrapolation. 

\smallskip

We can now construct an embedding of the group of dimension functions on $\Gamma$ 
as a subgroup of the infinite braid group $B_\infty$. We define the map as in \eqref{phifsigma},
with $\varphi(f) =\varphi(\lceil f \rceil)$, and with $v_j=0$ for each undefined vertex. 

\smallskip

To see that this is an embedding, it suffices to check the map on the generators of
$K_0(A)$. As shown in \cite{Kerov}, the dimension function corresponding to $1\in K_0(A)$ is 
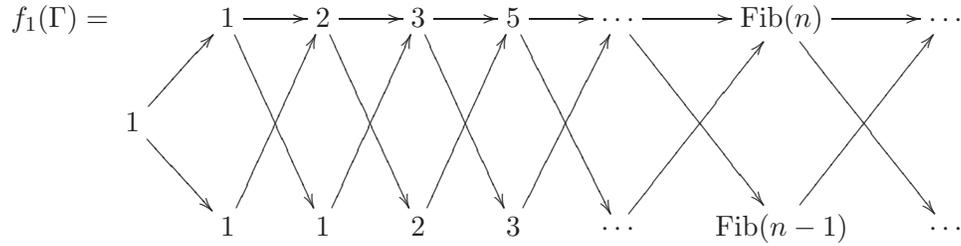
\[f_1(\Gamma)=\xymatrix{
& 1 \ar[r] \ar[rdd] & 2 \ar[r] \ar[rdd] & 3 \ar[r] \ar[rdd] & 5 \ar[r] \ar[rdd] & \cdots \ar[r]\ar[rdd] & \textrm{Fib}(n)\ar[r]\ar[rdd] & \cdots \\
1 \ar[ur] \ar[dr] \\
& 1 \ar[ruu]  & 1 \ar[ruu] & 2 \ar[ruu] & 3 \ar[ruu] & \cdots \ar[ruu] & \textrm{Fib}(n-1)\ar[ruu] & \cdots} \]
The corresponding element in $B_\infty$ is the product 
$\varphi(f_1)=\sigma=\prod_n \sigma_{2n+1}^{\textrm{Fib}(n)}$.

\smallskip

Using the expression for the generator $f_1$ and the action of the shift of the continued
fraction expansion, it can be deduced that one also has
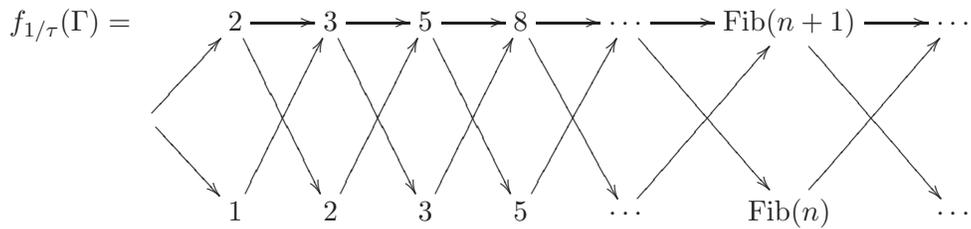
\[f_{1/\tau}(\Gamma)=\xymatrix{
& 2 \ar[r] \ar[rdd] & 3 \ar[r] \ar[rdd] & 5 \ar[r] \ar[rdd] & 8 \ar[r] \ar[rdd] & \cdots \ar[r]\ar[rdd] & \textrm{Fib}(n+1)\ar[r]\ar[rdd] & \cdots \\
{} \ar[ur] \ar[dr] \\
& 1 \ar[ruu]  & 2 \ar[ruu] & 3 \ar[ruu] & 5 \ar[ruu] & \cdots \ar[ruu] & \textrm{Fib}(n)\ar[ruu] & \cdots} \]
so that the dimension function corresponding to $\tau$ is
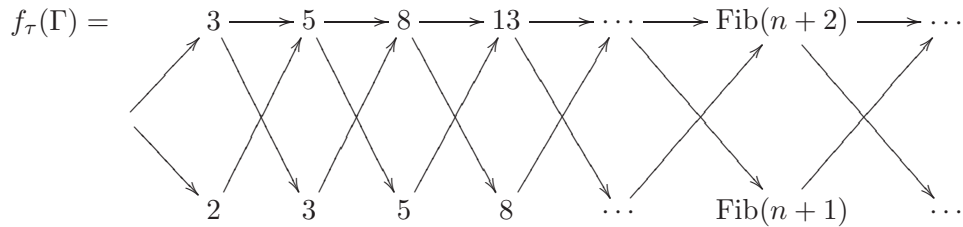
\[f_{\tau}(\Gamma)=\xymatrix{
& 3 \ar[r] \ar[rdd] & 5 \ar[r] \ar[rdd] & 8 \ar[r] \ar[rdd] & 13 \ar[r] \ar[rdd] & \cdots \ar[r]\ar[rdd] & \textrm{Fib}(n+2)\ar[r]\ar[rdd] & \cdots \\
{} \ar[ur] \ar[dr] \\
& 2 \ar[ruu]  & 3 \ar[ruu] & 5 \ar[ruu] & 8 \ar[ruu] & \cdots \ar[ruu] & \textrm{Fib}(n+1)\ar[ruu] & \cdots} \]

\smallskip

The subgroup of $B_\infty$ is then the abelian subgroup generated by $\varphi(f_1)$ and
$\varphi(f_\tau)$.
\endproof

\bigskip
\subsection{Additional questions}

It is known that the Fibonacci anyon system is universal, that is, that the braiding
of anyons suffices to approximate arbitrary unitary operators on the Hilbert space
of the system, see \cite{FreeKit}, \cite{NSSFS}, \cite{Pres}. It is therefore natural to ask whether
a similar universality property may hold for the anyon systems constructed in
Theorem \ref{RManyons} from quantum tori with real multiplication.

\smallskip

Quantum tori with real multiplication have additional structure, including
a homogeneous coordinate ring \cite{Polishchuk}, constructed using
the basic modules and holomorphic structures. This homogeneous ring
is also related to quantum theta functions \cite{Vlasenko}. Quantum 
theta function in turn have interesting relations to Gabor frames \cite{Luef}. 
It would be interesting to see if some of these additional structures on
real multiplication quantum tori also admit interpretations in terms
of anyon systems constructed using the basic modules.

\smallskip

Quantum tori also play a prominent role in the study of quantum Hall 
systems and AF algebras occur in the modeling of quasi crystals, 
\cite{Bell1}, \cite{Bell2}. In view of their respective role in the 
construction of anyon systems
described above, it would be interesting to seek concrete realizations
of the anyons in terms of some of the physical systems related to the
geometry quantum tori.

\bigskip

\subsection*{Acknowledgment} The second author is supported by
a Summer Undergraduate Research Fellowship at Caltech. 
The first author is supported by NSF grants
DMS-0901221, DMS-1007207, DMS-1201512, PHY-1205440.

\end{document}